\documentclass[twocolumn]{pasj00}
\begin{document}

\SetRunningHead{Improvement in calibration of GSO scintillators in the \textit{Suzaku} HXD}{S. Yamada et al.}
\title{Improvements in calibration of GSO scintillators \\ in the \textit{Suzaku} Hard X-ray Detector}

\author{%

Shin'ya \textsc{Yamada}\altaffilmark{1}, 
Kazuo \textsc{Makishima}\altaffilmark{1}, 
Kazuhiro \textsc{Nakazawa}\altaffilmark{1},
Motohide \textsc{Kokubun}\altaffilmark{2}, \\
Madoka \textsc{Kawaharada}\altaffilmark{2},
Takao \textsc{Kitaguchi}\altaffilmark{5}, 
Shin \textsc{Watanabe}\altaffilmark{2},\\
Hiromitsu \textsc{Takahashi}\altaffilmark{3},
Hirofumi \textsc{Noda}\altaffilmark{1},
Hiroyuki \textsc{Nishioka}\altaffilmark{1}, 
Kazuyoshi \textsc{Hiragi}\altaffilmark{3}, \\
Katsuhiro \textsc{Hayashi}\altaffilmark{3},
Kenta \textsc{Nakajima}\altaffilmark{1}, 
Makoto \textsc{Tashiro}\altaffilmark{4},
Makoto \textsc{Sasano}\altaffilmark{1}, \\
Sho \textsc{Nishino}\altaffilmark{3}, 
Shunsuke \textsc{Torii}\altaffilmark{1},
Soki \textsc{Sakurai}\altaffilmark{1},
Tadayuki \textsc{Takahashi}\altaffilmark{2,1}, \\
Tsunefumi \textsc{Mizuno}\altaffilmark{3}, 
Teruaki \textsc{Enoto}\altaffilmark{6}, 
Takayuki \textsc{Yuasa}\altaffilmark{1},
Takaaki \textsc{Tanaka}\altaffilmark{7}, \\
Tomomi \textsc{Kouzu}\altaffilmark{4},
Toshio \textsc{Nakano}\altaffilmark{1}, 
Yasushi \textsc{Fukazawa}\altaffilmark{3}, 
Yukikatsu \textsc{Terada}\altaffilmark{4}, \\
Yasunobu \textsc{Uchiyama}\altaffilmark{7},
Wataru \textsc{Iwakiri}\altaffilmark{4},
and the HXD team
}

\altaffiltext{1}{
   Department of Physics, University of Tokyo\\
   7-3-1, Hongo, Bunkyo-ku, Tokyo, 113-0033, Japan}
%\altaffiltext{2}{
%   Cosmic Radiation Laboratory, Institute of Physical and Chemical 
%   Research (RIKEN)\\
%   2-1 Hirosawa, Wako-shi, Saitama, 351-0198, Japan}
\altaffiltext{2}{Department of High Energy Astrophysics, Institute of 
Space and Astronomical Science (ISAS),\\
Japan Aerospace Exploration Agency (JAXA),
3-1-1 Yoshinodai, Sagamihara, Kanagawa 229-8510}
\altaffiltext{3}{Department of Physics, Hiroshima University,\\
1-3-1 Kagamiyama, Higashi-Hiroshima, Hiroshima 739-8526}
\altaffiltext{4}{Department of Physics, Saitama University, \\
Shimo-Okubo, Sakura-ku, Saitama-shi, Saitama 338-8570}
\altaffiltext{5}{
California Institute of Technology, 
1200 East California Blvd., 
Pasadena CA 91125, USA
}
\altaffiltext{6}{
KIPAC/Stanford University
2575 Sand Hill Road, 
Menlo Park, CA 94025, USA
}
\altaffiltext{7}
{Stanford Linear Accelerator Center, 
2575 Sand Hill Road, 
Menlo Park, CA 94025, USA}

\email{yamada@juno.phys.s.u-tokyo.ac.jp}


\Received{2010 November 14}
\Accepted{2011 May 23}
\Published{2011 October 25}

\maketitle

%%%%% Abstract %%%%%
\begin{abstract}

Improvements of in-orbit calibration of GSO scintillators 
in the Hard X-ray Detector on board Suzaku are reported. 
To resolve an apparent change of the energy scale of GSO 
which appeared across the launch for unknown reasons, 
consistent and thorough re-analyses 
of both pre-launch and in-orbit data have been performed. 
With laboratory experiments using spare hardware, 
the pulse height offset, corresponding to zero energy input, 
was found to change by $\sim 0.5$\% of the full analog voltage scale, 
depending on the power supply. 
Furthermore, by carefully calculating all the light outputs 
of secondaries from activation lines used in the in-orbit gain determination, 
their energy deposits in GSO were found to be effectively lower, 
by several percent, than their nominal energies.
Taking both these effects into account, 
%the in-orbit data were confirmed to be consistent with the on-ground measurements. 
the in-orbit data agrees with the on-ground measurements within $\sim $5\%, 
without employing the artificial correction introduced in the previous work (Kokubun et al. 2007). 
%With this knowledge, the GSO data processing and the response matrix of GSO were updated. 
%As a result, the HXD-PIN and HXD-GSO spectra of the Crab Nebula are
%now expressed successfully over 12-300 keV by a broken powerlaw 
%with a break energy of $\sim 110$ keV. 
With this knowledge, we updated the data processing, the response, and the auxiliary files of GSO, 
and reproduced the HXD-PIN and HXD-GSO spectra of the Crab Nebula 
over 12-300 keV by a broken powerlaw with a break energy of $\sim 110$ keV. 

\end{abstract}

%%%%% Intro %%%%%
%==================
% S.1
\section{Introduction}
\label{section:1}

%\green{ひじょうに簡単なすざくとHXD紹介}
The fifth Japanese X-ray satellite Suzaku (\cite{Mitsuda2007}), carries the X-ray Imaging Spectrometer (XIS; \cite{Koyama2007}) located at the foci of the X-ray Telescope (XRT \cite{Serlemitsos2006}) 
and a non-imaging hard X-ray instrument, the Hard X-ray Detector (HXD). 
The HXD covers a hard X-ray energy range of 10--600 keV, utilizing Si-PIN 
diodes (hereafter HXD-PIN), and gadolinium silicate scintillators (Gd$_2$SiO$_5$:Ce, hereafter HXD-GSO) 
which are placed behind HXD-PIN. 
The detailed design of the HXD is summarized in Takahashi et al. (2007) (hereafter Paper I), 
followed by a report on its in-orbit performance (\cite{Kokubun2007}; hereafter Paper II) 
and that of the timing property \citep{Terada2008}. 

%\green{地上でも上空でもcalした}
Prior to the launch on 2005 July 10, 
the performance of the HXD was tested and calibrated at every stage of its integration, and 
also after the instrument was mounted on the spacecraft. 
In addition to these pre-launch experiments, 
several Monte-Carlo simulations were performed, 
 to better understand the HXD performance (including its background) expected in near-Earth 
radiation environments (\cite{Terada2005}; Paper I). 
We have also calibrated the HXD performance extensively after the launch (Paper II), and  
compared all propertied that can be obtained in-orbit with the results of the ground measurements. 
As a result, we have successfully verified that the HXD performs in-orbit generally as expected (Paper II), 
and reconfirmed most of the results of the pre-launch calibration. 

%\green{その結果、矛盾が残っており、その解決が目的}

In spite of these successful in-orbit calibration efforts, 
we are still left with a few unsolved issues of the HXD performance. 
One of them is an apparent change in
the energy scale (relation between the incident X-ray energy and detector pulse height) of HXD-GSO.  
We found this issue when we observed the X-ray pulsar A0535+26, 
which exhibits an absorption line at $\sim 50$ keV \citep{Terada2006}; 
this spectral feature appeared at discrepant energies in the HXD-PIN and HXD-GSO 
spectra that were produced based on the nominal pre-launch calibration (see subsection 2.3). 
Since the energy scale of HXD-PIN was kept unchanged across the launch 
(figure 5 of Paper II), 
we tentatively concluded that the energy scale of HXD-GSO changed for some unspecified reasons, and hence 
added an artificial non-linearity to the GSO energy scale (figure 12 of Paper II) as a 
temporary solution. 

The aim of the present paper is to more fundamentally solve this issue, by revisiting the pre-launch and in-orbit calibrations, 
and conduct some laboratory experiments using spare hardware. 
Section 2 is used to review the way of conversion from energy to pulse height, 
and to present the investigation strategy. 
In section 3, we study the spare hardware performance, 
followed in section 4 by a detailed comparison between the on-ground and in-orbit data, 
and construction of a revised energy scale. 
We present in section 5 how well our new solution worked on actual observation data, and 
summarized these works in section 6. 
 
%==================
% S.2
%   
\section{The HXD Experiment}

We start with a brief summary of the HXD in section 2.1, 
and then the procedure of the energy scale in section 2.2. 
In section 2.3, 
the in-orbit calibration that 
had been performed until Paper II are summarized. 
Other unsolved issues are introduced in section 2.4. 
We present our methods to tackle the problem in section 2.5.

\label{section:2}

\begin{figure*}[htbp]
    \begin{center}
      \FigureFile(140mm,140mm){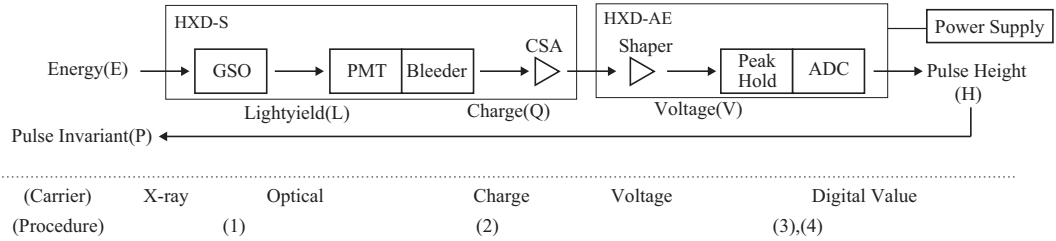}
%      \FigureFile(140mm,140mm){figure1.pdf}
%      \includegraphics[width=0.9\textwidth]{figure1.pdf}	
            \end{center}
    \caption{A functional diagram of the energy scale of GSO. The signal carrier at each stage and the corresponding steps listed in subsection 2.2 are summarized at the bottom.}
    \label{ped}
\end{figure*}

\subsection{The hardware configuration}

As detailed in Paper I, the HXD consists of three major components:
a sensor part (HXD-S), an analog electronics part (HXD-AE), and a digital electronics system (HXD-DE). 
HXD-S has a compound-eye configuration, made of 4$\times$4 well-type phoswich detectors 
(``Well-counter units''), surrounded by 20 thick active shields (``Anti-counter units''). 
A Well-counter unit consists of 4 Si-PIN diodes and 4 GSO scintillators, all placed inside a deep well 
of active shields formed by BGO scintillators. 
Each unit outputs 4 Si-PIN signals and one GSO signal, 
after scintillation light from the 4 GSOs 
is detected by a single photomultiplier tube (PMT) which also detects light from the BGO shield. 
All signals of 36 PMTs (16 Well-counter and 20 Anti-counter units) and 64 PIN diodes are fed in parallel into HXD-AE, where they are amplified, 
pulse-shape discriminated, and analog-to-digital converted. 
The digitized signals are further processed by HXD-DE using an on board CPU. 

The present paper pertains to the GSO scintillators (HXD-GSO) of the Well-counter units, as 
well as to HXD-AE, but not to HXD-DE.
It does not deal with the Anti-counter units in HXD-S, either. 

\subsection{Energy to pulse-height conversion}

% EがどのようにしてPHになるか、賢いブロック図などを
% 援用し、徹底的にわかり易く述べる。ここがポイントの１つ。
Let us briefly review the flow of converting the pulse-height into energy as described in Paper II. 
In any radiation detector that detects individual particles 
(including photons) as electronic pulses, 
the energy scale is critically important to its performance.
Its determination is significantly more difficult in hard X-rays, 
than in energies below $\sim 25$ keV where 
we can utilize various celestial/instrumental atomic lines and built-in calibration radio-isotopes. 
In the particular case of HXD-GSO, 
the energy $E$ of a detected photon is converted into pulse height, 
or more properly, ``pulse invariant'', $P$, through a process shown in figure 1. 
A photon with an energy of $E$, totally absorbed in a GSO scintillator,  
creates a photoelectron which immediately produces many secondary electrons. 
These electrons deposit their energies in the scintillator, creating optical photons
of a total number $L$ (which is called light yeild). 
The photons are (partially) detected by a PMT (Hamamatsu R6231),
 to get multiplied into an electronic pulse with charge $Q$. 
This charge is converted into a voltage pulse $V$ in a charge sensitive amplifier
(CSA). 
The process up to this stage takes place in HXD-S. 
The pulse sent to HXD-AE is further amplified and shaped 
therein  by a shaper,  then peak-held in an application-specific integration circuit, 
and finally digitalized by a 12-bit analog-to-digital converter. 
Thus, a photon energy up to $E \sim 600$ keV is converted into a 12-bit pulse height, $H$. 

Although the pulse height $H$ is approximately proportional to $E$ with a constant coefficient of 
proportionality, there are at least following effects that can cause deviations and/or
time variations in the $E$ vs. $H$ proportionality.

\begin{enumerate}
\item[(1)] Non-linearity in the GSO light yeild \citep{Uchiyama2001}. 
\item[(2)] The overall coefficient of proportionality, or ``gain'', donated  $G$.
\item[(3)] Differential and integrated non-linearity in the amplification chain in HXD-AE \citep{Kitaguchi2006}.  
\item[(4)] The offset signal in analog electronics, or the ``pedestal'' defined in Paper II, denoted $H_{\rm{o}}$ in this peper. This means a value of $H$ when no signal is present in GSO ($E=L=Q=0$). 
\end{enumerate}

The effect (2) is further decomposed into the GSO light yield, the PMT gain, 
the CSA capacitance, and electronics settings. 
As detailed in Paper II, 
the GSO light yield depends on the temperature, while the PMT gain varies by $\sim 10$\% 
in a complex way, depending on the temperature, time since the launch, and time after passing through South Atlantic Anomaly, wherein the PMT high voltage is turned off for $\sim 20$ minutes or less. 
Therefore, this coefficient, $G$, must be determined from moment to moment. 
In contrast, the effects (1), (3), and (4) are considered approximately constant with time, and were 
determined in the ground measurements; therefore, 
we can apply these results to the in-orbit data unless something in hardware did change. 

Given the knowledge on the effects (1)-(4), 
we can convert the pulse height $H$ of each event into a quantity called pulse invariant ($P$) 
through the reverse procedure of figure 1.  
If (1)--(4) are correct, $P$ should coincide with $E$ within the detection energy resolution (statistical effect) and calibration uncertainties in the energy scale (systematic effects).

\subsection{In-orbit calibration until Paper II}

\begin{figure*}[htbp]
    \begin{center}
%      \FigureFile(140mm,140mm){figure2.pdf}
      \FigureFile(100mm,100mm){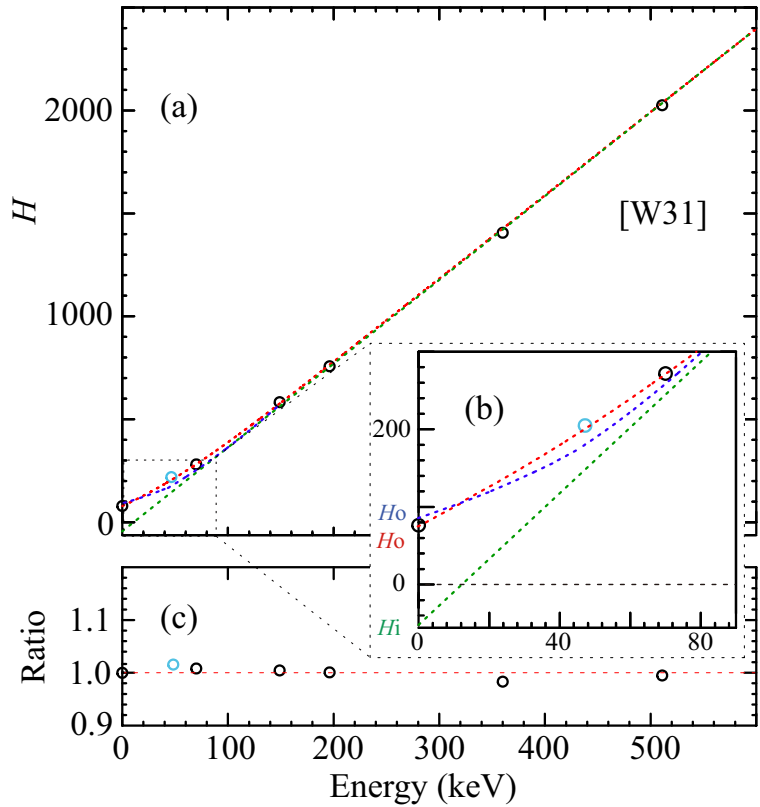}
    \caption{ (a) Initial in-orbit calibration of the GSO energy scale. 
Abscissa is the input energy $E$, while ordinate is the pulse height $H$. In addition to several background lines, the cyclotron line of A0535+26 at $\sim 50$ keV (cyan), and the``pedestal" information at 0 keV are also shown. The green line represents the linear relation derived by fitting three background lines, 
    and the red one is the final energy scale of GSO derived in the present study. 
    The blue curve is the tentative solution introduced in Paper II by invoking an artificial bending. 
    (b) An expanded view of the lower energy range of (a). 
    (c) The ratio of the red line to the data points.}
       \label{ehplot}
    \end{center}
\end{figure*}

% 切片が -50 から -10 に変わった
During the initial in-orbit calibration phase of Suzaku, 
we applied the ground-determined (1), (3), and (4) to the actual in-orbit HXD-GSO data. 
As explained in section 2.2, $G$ must be determined every $\sim 50$ minutes, 
so we used for that purpose GSO background lines at 511 keV, 
$\sim$350 keV, and 153 keV (table. 7 of Paper II for details). 
At that time, 
the energies of these lines were considered to be well defined, 
because they are based on nuclear physics, 
and the effect of non-linearity of (1) and (3) are negligible at least in $> 100 $ keV.  
Thus, by fitting these lines on the $E$ vs. $H$ plane with a linear function [equation (1) in Paper II], 
we determined the slope $G$ and the intercept of the line with the $H$ axis, 
hereafter $H_{\rm{i}}$, allowing both to vary.
As a typical case,
the employed data points (on the $E$ vs. $H$ plane) and the linear function (green) 
for one Well Unit (W31) are shown in figure \ref{ehplot},
which is almost the same as figure 12 in Paper II. 
The data points include background gamma-ray lines at 70 keV, 150 keV, 193 keV, $\sim$ 350 keV, and 511 keV. 
Here $H$ is expressed in units of ADC channels, with its maximum, 4096, corresponding to $\sim$ 700 keV. 
Then, $H_{\rm{i}}$ has appeared at $\sim -50$ channels (Paper II), 
in contrast to $H_{\rm{i}} \sim -10$ channels measured in the ground calibration (Paper I).  
This means apparent changes by $\Delta H_{\rm{i}} = -40$ between the post- and pre-launch calibrations. 

\begin{figure*}[htbp]
    \begin{center}
%      \FigureFile(140mm,140mm){figure3.pdf}
      \FigureFile(100mm,100mm){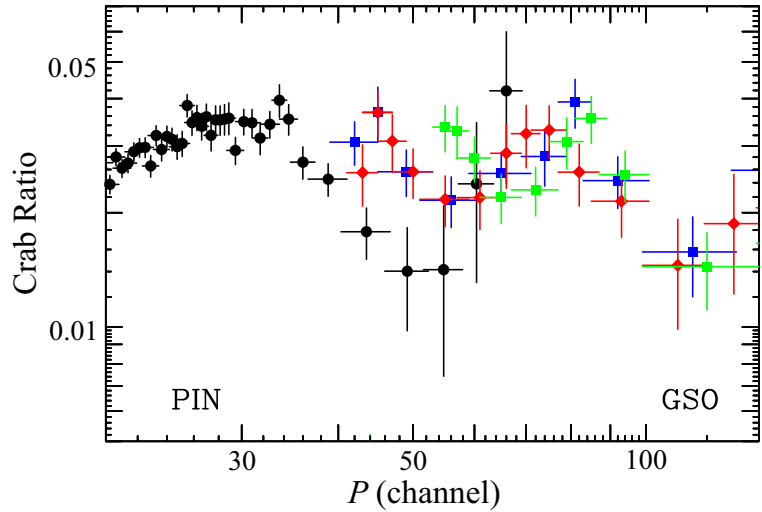}
            \end{center}
    \caption{Crab ratio spectra of A0535+26, as a function of the pulse height invariant, $P$. 
    The one obtained with HXD-PIN is shown in black. 
    The green one is obtained with HXD-GSO with the initial energy scale, 
    while the blue one is after including the artificial correction given in figure 2 (blue curve) and Paper II.
    The red one is created by using the energy scale obtained in this paper.}
    % pwd
    % /net/work/yamada/ana/suzaku/hxdcal/KijyouData/All_Data/work_ver2.0/ver2.0/100021010/hxd/event_uf_default/CrabRatio
    % Paper_2ndref_A0535_ratio_crab_a0535ver0.5_1.top より               
    \label{a0535}
\end{figure*}

% サイクロトロン吸収が 60 keV に見えた
Although the non-zero value of $\Delta H_{i}$ suggests some changes in the GSO energy scale in energies of
 $\lesssim 70 $ keV, there are no appropriate spectral features in the $\sim$ 10 to $\sim$ 70 keV range to be used for calibration. 
Accordingly, a celestial object named A0535+26 was utilized to calibrate the energy scale down to $\sim$ 50 keV \citep{Terada2006}, 
because it is known to exhibit a cyclotron resonance absorption line as shown in black in figure \ref{a0535}. 
The feature was clearly detected by HXD-PIN at an energy of $\sim 45$ keV (\cite{Terada2008}). 
However, as shown in green in figure \ref{a0535} and plotted in cyan in figure \ref{ehplot}, 
the same feature was seen at about 60 keV ($H \sim 205$) in the GSO spectrum, 
when the detected $H$ was converted  back to $P$ by applying (1)-(4) and 
extrapolating the energy scale determined over 70--511 keV 
(doted green line in figure 2)
to energies below 70 keV. 
More specifically, this procedure appears to assign too high a value of $E$ for a given $H$ of HXD-GSO, 
 or conversely, to underestimate $H$ for GSO events with $E \sim 45$ keV. 
This could be regarded as being caused by the change of $\Delta H_{\rm{i}} \simeq -40$ mentioned above.  

% 問題は HXD-PINではない。
The observation of A0525+26 thus revealed a $\sim$ 20\% discrepancy 
between the energy scales of HXD-PIN and HXD-GSO at lower energies. 
However, the energy calibration of HXD-PIN is more reliable than that of GSO, because its energy scale 
is mostly determined by the preamplifier capacitance.
Furthermore, an instrumental peak due to escaping Gd-K line photons, 
 expected at 43 keV, correctly  appeared at $43\pm1$ keV in the in-orbit background spectra of HXD-PIN. 
Therefore, we had to conclude that the problem is in HXD-GSO, rather than in HXD-PIN. 

% ここで初めて「pedestal」を、パラグラフを改めて、きっちり述べる。
In order to reconfirm the inference derived from the A0535+26 data, 
Paper II utilized pedestal information $H_{\rm{o}}$ on the PMT signal, i.e., the value of $H$ 
when the energy deposit $E$ approaches 0. 
This $H_{\rm{o}}$ can be obtained not only during the pre-launch tests, but also 
from the in-orbit data, by referring to values of $H$ of those events which were triggered by the PIN diodes in the same 
Well-counter unit: even for such events, the PMT pulse heights are latched, and sent out to telemetry, 
for later use in the background rejection. 
In fact, this gives $H_{\rm{o}} + \delta $, where $\delta$ is contribution from noise in the PMT signal and 
can be estimated to be less than $\sim $1 keV ( $\sim$ 4 ch).  The value of $H_{\rm{o}}$ derived in this way is given in figure 2b as a data point at $E \sim 0 $ and $H_{\rm{o}} \sim 80$. 
Thus, the problem uncovered by the A0535+26 data was reconfirmed. 
This effect exceeds what can be explained by (1) and (3). 

% Paper IIの第２項を入れて、人工的に合わせ込んだ。。
To temporarily solve this apparent discrepancy in the energy scale of GSO below $\sim 70$ keV, 
an additional non-linearity (the third term of equation 2 in Paper II) has been artificially introduced. 
As a result, the GSO spectrum of A0535+26 moved to lower energies (blue spectrum in figure \ref{a0535}), 
and the cyclotron absorption-line energy measured with HXD-GSO became consistent with that of HXD-PIN. 
Although the blue spectrum in figure \ref{a0535} still shows a somewhat higher absorption-line energy 
than the black one, the difference can be explained away by 
a poorer energy resolution of HXD-GSO than HXD-PIN, coupled with a steeply declining continuum. 
In figure \ref{a0535}, the entire spectra are divided by the Crab Nebula spectrum 
which behaves as $\propto E^{-2.1}$.

%\subsection{Other differences from on-ground data before Paper II}
\subsection{Other differences between on ground and in-orbit data}

In addition to the above issue of the GSO energy scale, the in-orbit data exhibited 
some other anomalies which cannot be fully explained by the pre-launch calibration. 
%At that time,
When Paper II was complied, 
it was not clear 
whether either or both of these two unexpected effects comes from the same origin as the GSO energy scale problem. 
These issues are discussed in section 5. 

One is the performance of pulse-shape discrimination, which is applied to the PMT signal in HXD-AE to distinguish 
GSO and BGO events (see figure 15 in Paper II).  
As described later in subsection 5.1, width of the GSO branch on so-called slow-fast diagram, 
appears to have charged across the launch. 
The other is that the current response matrix for GSO is overpredicting the spectrum by 10\% 
in energies below 100 keV, as evidenced by a Crab spectrum of GSO (figure 18 in Paper II). 
For a temporary settlement, an artificial ancillary response file has been 
provided.

\subsection{Working hypotheses and verification methods}
Although the unexpected change in the GSO energy scale below $\sim 70$ keV is at present absorbed 
by the artificial correction factor, this is a temporary and empirical solution. 
More fundamentally, we need to understand the cause of the change, and if necessary, update the GSO energy scale.

In the following, 
we conduct two approaches towards the above objectives. 
One of them is to  examine the HXD hardware performance 
for any launch-related changes.
This is carried out in section 3,
partially incorporating laboratory experiments using scintillators, 
phototubes, preamplifiers, and analog electronics 
that are all (nearly) equivalent to the actual flight-model HXD. 

The other approach to our goal is to 
examine the process of energy calibration, 
searching for any inaccuracy or inconsistency
between the pre- and post-launch calibrations. 
In section 4, we therefore re-analyze the in-orbit GSO data, 
as well as the pre-launch calibration data  
which were taken by irradiating various calibration isotopes to the HXD. 
Through a consistent data analysis, 
we try to understand more accurately
how the energy scale had changed across the launch.

%==================
% S.3
\section{Examination of the hardware performance}
\label{section:3}

%(ここを前文として扱う) 

\subsection{Possible changes in the HXD hardware}

The first of the two approaches mentioned in section 2.5 is based on a concern 
that the present issue can be a result of some changes 
in the HXD hardware performance,
including its scintillators, PMTs, CSAs, bleeders, and electronics; 
in other words, anywhere in the diagram of figure 1. 
This is further subdivided into 
the following three general possibilities. 
One is that the launch stress caused some irreversible changes (including damages)
 to the HXD hardware. 
Another is that frequent large-amplitude scintillation pulses, 
produced by charged particles in the space environment, 
brought about some differences in the low-energy response.
The other is that the spacecraft power supply, 
which is known to be less stable  than those used in laboratory, 
may have caused some changes in the electronics response to low-pulse-height signals.

We found that the former two possibilities do not affect the energy scale, 
so that we summarized them in Appendix 1 and 2, respectively.  
In the next subsection, we examine the third possibilities that 
actually affected the energy scale of the GSO. 

\subsection{Effects of the power supply}
\label{subsection:3-}

A clear difference in the hardware environment between the ground and in-orbit measurements is 
the power supply. Our ground calibration was performed using general-purpose power supplies 
(+12 V, -12 V, +5 V analog, and +5 digital), which  
are more stable than the flight one. 
This could affect the HXD-AE performance, particularly at low pulse heights. 
Although HXD-AE does not have its own calibrators, 
the raw pulse height accumulated during no signal inputs in GSO, i.e., pedestal ($H_{\rm{o}}$ in figure \ref{ehplot}), can be used for this purpose. 
The values of the pedestal have been compared among three data sets: 
(1) the on-ground calibration data acquired in June 2004, using a laboratory power supply, 
where the HXD was not yet mounted on the spacecraft;  
(2) the on-ground data taken in 2004 August with the HXD mounted on the spacecraft, 
and driven by the flight power supply; and 
(3) in-orbit data acquired from a black sky in June 2007 
during a period when the PMT high voltage was temporary turned off 
to avoid over/under shoot caused by cosmic ray signals. 
As explained in section 2.3, 
the pedestal signal, $H_{\rm{o}}$, was obtained 
by selecting cosmic rays or background events triggered uniquely by HXD-PIN. 

Figure \ref{ped} shows histograms of $H_{\rm{o}}$ for the 16 Well-counter units. 
Thus, they show $H_{\rm{o}} \sim 81$ channels when 
HXD-AE was driven by laboratory power supplies. 
In contrast, they exhibited $H_{\rm{o}} \sim 73$ channels when the flight power supply was used, 
both before or after the launch. 
The change by $\Delta H_{\rm{o}} \sim -8$ can be attributed to the difference in the power supplies, 
because everything else was the same. 
However, as illustrated in figure \ref{ehplot}, the effect, 
at most $\sim$4\% for the 50 keV signals (equivalent to $H \sim 200$), 
is not large enough to account for the $\sim 20$\% change of energy scale 
at $\sim$ 50 keV (section 2.3); 
some other effects are needed. 

\begin{figure}[htbp]
    \begin{center}
%      \FigureFile(140mm,140mm){figure5.pdf}
      \FigureFile(70mm,70mm){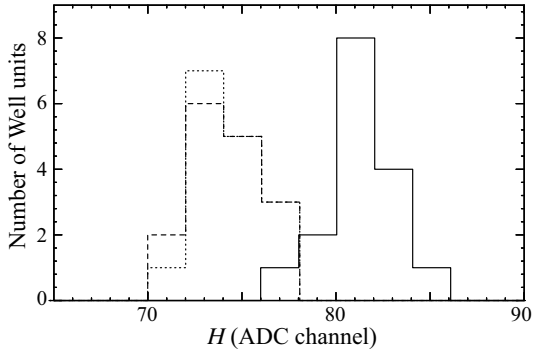}
            \end{center}
    \caption{Histograms of the number of Well-counter units as a function of $H_{\rm{o}}$ for the three conditions; 
    the solid, dotted, and dashed lines indicate pre-launch data taken in laboratory, those obtained on the spacecraft, and in-orbit data, respectively.}
    \label{ped}
\end{figure}

%------------------------------------------  Section 4 
\section{A Unified Analysis of Ground and In-Orbit data sets}
\label{section:3}

We start with examination of ground calibration data in section 4.1, 
and then, employing newly introduced ``$Q$ vs. $H$" plot in section 4.2,  
compare the ground data to in-orbit data in section 4.3. 
The examination on light yield and reanalysis of in-orbit data are 
summarized in section 4.4 and 4.5.

\subsection{Summary of ground calibrations}

%・以下のパラグラフに対する大まかな流れを提示。
Since we have found that the change in hardware cannot explain the change of the $E$ vs. $H$ relation, 
we then proceed to the second approach introduced in section 2.5, and 
try to review our calibration processes, starting with that on ground data. 

%・GSOの発光非線形性 (E vs. L)を、ガンマ線アイソトープと、
%　an ideal ADC で較正したこと (Kitaguchi 2006か？)。これが
%　E vs. H の非線形性の主因を担っていること。

Although an inorganic scintillator has a roughly constant $L$ (subsection 2.2) of its own, 
the $E$ vs. $L$ proportionality is in many cases only approximate, and needs detailed calibration. 
such measurements of GSO used in the HXD were performed by Kitaguchi et al. (2005), 
irradiating gamma-rays from various radio-isotopes. 

%・テストパルスで、HXD-AE の Q vs. H 関係を較正したこと
%    (Kawaharada 2004). 大きな非線形性は無い。
$Q$ vs. $H$ calibration on ground in June 2004 were performed by  Kawaharada et al. (2004), 
by inputting electronic test pulses of various pulse heights simultaneously to HXD-AE and a laboratory MCA (the same as used in section 3.2). 
When the latter was used as an ideal reference, 
HXD-AE was found to have a good $Q$ vs. $H$ proportionality at $H \gtrsim 200$ ($\gtrsim$50 keV), 
with slight non-linearity below $H \sim 200$. 

%・One of the 16 Well counter units と、the flight HXD-AEを
%　用い、end-to-end テストを行い、軌道上とほぼ同じ配置での
%　データを得たこと。電源だけが異なること。これを以下の比較
%　に用いることを、クリアに宣言すべし。
%・end-to-end calの結果は、上記２つの個別calを組み合わせて
%　予想されるのと、矛盾しなこと(ですよね？)。
In addition to these basic measurements, 
end-to-end pre-launch calibrations were performed by irradiating gamma-ray isotopes 
to GSO from outside and processing the signals with HXD-AE. 
The acquired data are of high importance, 
because they can be compared directly with the in-orbit data. 

Here, we have analyzed such pre-launch GSO calibration data, taken by Kawaharada et al. (2004) in June 2004 
under the same condition as (1) in section 3.2 
(i.e., using a laboratory power supply). 
Of the 16 Well-counter units, we use the W31 unit for simplicity, 
because the HXD-AE performance is almost the same among the 16 channels (Paper I). 
Thus, we can use the results to compare them with the in-orbit data.

\subsection{Introduction of  ``Q vs H'' plots}
%(あえて単独のサブセクションにしたが、これにこだわらない）
%## なぜこのプロットが良いか、じっくり説明すべし。
When comparing different sets of calibration data,
we usually create from each data set an ``$E$ vs. $H$ plot",
wherein the measured pulse heights $H$ are
plotted against the known photon energies $E$.
A comparison between a pair of $E$ vs. $H$ plots requires
their appropriate corrections for the difference
in the overall gain $G$,
which varies depending mostly on the PMT gain. 
This is usually done by renormalizing the $H$ values 
from one of the plots to a certain constant factor,
so that these plots become close to each other. 
This renormalization procedure is correct if, for example,
the $E$ vs. $H$ relation deviates from an exact proportionality
mainly due to the GSO non-linearity between $E$ and $L$; 
however, it is not if the deviation is dominated by electronics effects
and hence is dependent on $H$, not $E$ or $L$. 
More generally, an $E$ vs. $H$ relation may be approximated as $H= GE+H_{\rm{o}}$ at small values of $E$,
and it is not trivial how to deal with $H_{\rm{o}}$ when re-normalizing; 
should we calculate as $H \rightarrow \beta(GE+H_{\rm{o}})$, or $\beta GE+H_{\rm{o}}$, 
or in other ways (where $\beta$ is the renormalization factor)?

To avoid the above problem,
we have introduced a new concept of ``$Q$ vs. $H$" plot (Yamada 2008),
where $Q$ means the charge (figure 1) coming out of the PMT.
The new plot can be obtained from an ordinary
$E$ vs. $H$ plot in the following manner. 
First, the energy $E$ (assumed to be known) of each data point
is converted into the light yield $L$ through the knowledge on the GSO light yeild (Kitaguchi 2006), or $E$ vs. $L$ relation, which is implicitly assumed to be rather stable across the launch.
Then, the charge $Q$ is defined as a quantity proportional to $L$, 
without any zero-point offsets; this is justified by the general property of PMTs. 
The constant of proportionality between $Q$ and $L$
is determined so that the value of $Q$ becomes identical to that of $E$ (in units of keV) 
at the most reliable high-energy data point.
Thus, $Q$ is defined to take roughly the same values as $E$, 
and is considered to be free from the GSO non-linearity
and the PMT gain changes. 
The most time-variable factor, namely $G$, is re-normalized at this stage. 
In other words, we compare different calibration plots
by adjusting their energy axes, 
rather than by stretching/compressing their pulse-height axes. 
This procedure implicitly assume little changes in the HXD-AE performance.

\subsection{Comparison between the ground and in-orbit calibration results}

\begin{figure}[tbhp]
    \begin{center}
%      \FigureFile(140mm,140mm){figure6.pdf}
      \FigureFile(70mm,70mm){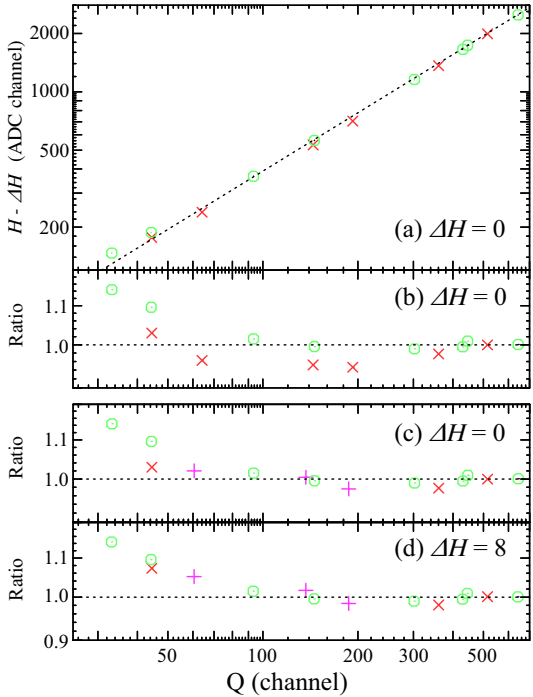}
    \caption{A $Q$ vs. $H$ plot. 
Abscissa is $Q$, while ordinate is $H$ but $\Delta H$ has been subtracted if needed. 
The gain is re-adjusted so that the on-ground data match the in-orbit ones. 
The errors are smaller than the size of plotting symbols. 
(a) The green circles are the pre-launch data, while the red crosses represent the in-orbit data points. 
The dotted black line is a proportionality line connecting the origin to the 511 keV point. 
(b) Ratios of the in-orbit and pre-launch data to the common line of proportionality defined in panel a. 
(c) The same as (b), but the value of $Q$ of the in-orbit lines at 70, 150, and 196 keV  (pink crosses) are re-calculated (section 4.4).  
(d) The same as (c), but $H$ of the in-orbit data are lowered by $\Delta H = 8$ ch 
considering the results of section 3.3}
    \label{newplot}
     \end{center}
\end{figure}

% 地上のデータ
Figure \ref{newplot} is the ``$Q$ vs. $H$'' plot constructed via the above procedure
from the ground calibration data (green) and in-orbit data (red).
The gain of the ground data were re-adjusted (along the horizontal axis)
so that they match the in-orbit ones. 
To guide the eyes, a proportionality line pointing from the origin to the data corresponding to 511 keV is superposed, 
and ratios of the data to this line are shown in figure \ref{newplot}b.  
In the lower-ADC channel ($\lesssim$ 200 ch) region, 
the ground calibration data exhibit $\sim10$\% 
deviation above the linear function.  
As described in subsection 3.3, 
this offset is attributable to the non-ideal $Q$ vs. $H$ properties of HXD-AE (Kawaharada et al. 2004). 
Actually, when the on-ground data points are corrected for this effect, 
the deviation from the proportionality line diminishes to less than $4$\%; 
this means that the energy scale of GSO for the on-ground data consists of all known factors. 

%上空データの検証}
Figure \ref{newplot}b reveals two systematic differences between
the pre-launch and in-orbit calibration data sets.
One is that the in-orbit data points fall systematically below the line of proportionality,
in energies from 60 to 400 keV:
this effect is not observed in the ground calibration data.
The other is that the in-orbit data points in the pulse-height range below $H \sim 250$ ch exhibit
much milder excess above the proportionality line, than the ground data points. 
Thus, the energy scale appears to have really changed across the launch. 
However, it is apparently inconsistent 
with the result that in-orbit data exhibit {\it larger} excess toward lower pulse heights than the pre-launch data 
(cf. figure 2; figure 12 in Paper II). 

% このプロットでの追加非線形性の解釈
As explained in section 2.2 and figure \ref{ehplot}, 
the first approximation of energy scale was derived 
by connecting the 150 keV, 350 keV, and 511 keV data points 
with a linear (not proportional) function, 
in which the slope and offset are both allowed to change. 
This linear function, with an intercept of $H_{\rm{i}} = -50$ (figure \ref{ehplot}), 
has a steeper slope than the line of proportionality (with $H_{\rm{i}} = 0$) employed in figure \ref{ehplot}.  
If this steeper line is instead used in figure \ref{newplot}, the in-orbit data in panel (b) would exhibit much larger 
positive residuals in $H \lesssim 200$, 
while the intermediate-energy data points would fall right on the line,
thus reproducing the behavior seen in figure \ref{ehplot}. 
Considering the result in section 3 that the hardware performance has not changed as much, 
it is therefore more reasonable to consider that the significant low-enegy excess observed in figure \ref{ehplot}
and Paper II is caused artificially 
because we started from an inappropriate linear function with a steeper slope 
(and hence with a large negative $H_{\rm{i}}$). 
If this is true, the next question is what has caused the line, 
connecting 150, 350, and 511 keV points, to have a wrong slope. 
Let us consider how the energies of these activation/annihilation lines seen in the in-orbit 
background spectra are measured with HXD-GSO. 

% ~/work/ana/suzaku/hxdcal/gain_KijyoOpe/all_hbk% cd testw31/Ly_matome/Mrom_final_fig/

\subsection{Re-examination of Light yield from activation lines}
% 光量計算の困難
Among the instrumental background lines used in the in-orbit energy calibration of HXD-GSO, 
the 511 keV electron-positron annihilation line is based on a simple elementary process, 
with its photo-peak created by absorption of a single mono-energetic gamma-ray line. 
In contrast, as described in Table 7 in Paper II, 
the activation lines at 70 keV, 150 keV, and 196 keV are 
statistical mixtures of various radiation components,
including nuclear transition gamma-rays, fluorescent X-rays, Auger electrons, and internal conversion electrons. 
This raise a suspect that the values of $L$ for the 150 keV and 350 keV activation lines 
would differ from their nominal values.

\begin{figure}[tbhp]
    \begin{center}
%      \FigureFile(140mm,140mm){figure7.pdf}
      \FigureFile(70mm,70mm){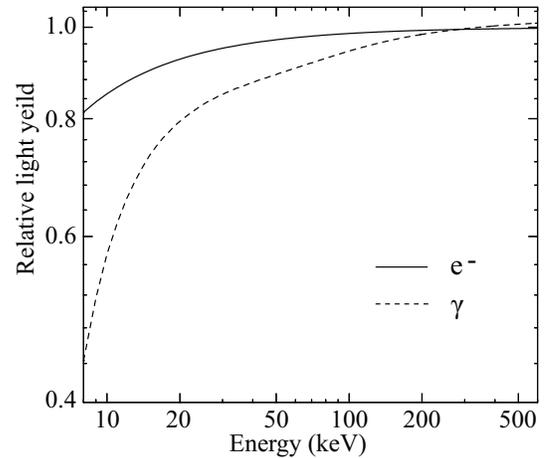}
          \end{center}
    \caption{Relative light yield of GSO as a function of energy. 
    The solid curve is for incident electrons (Uchiyama et al. 2002), 
    and the dashed one for photons (Kitaguchi et al. 2005).}
    \label{relaL}
\end{figure}
%~/work/ana/suzaku/hxdcal/gain_KijyoOpe/all_hbk/testw31/Calc_Mron_Ly

% これまで計算方法と問題点

The values of light yield, $L$, from activation lines have been so far calculated 
by assuming that their energy is deposited in GSO by a single gamma-ray photon. 
However, this approximation might be over-simplified, 
because electrons and photons give different $L$ in GSO and their energy deposited in GSO 
are not perfectly proportional to $L$.
Figure \ref{relaL} shows the light yield against energy 
for electrons (Uchiyama et al. 2001) and photons (Kitaguchi et al. 2006), 
which are both experimentally measured and normalized at 344 keV.  
Thus, the two curves are both considerably non-linear, and are different from each other. 
Therefore, even if the overall energy released by a nuclear decay is the same, 
the light output depends on 
how the total energy is split into individual components, 
and whether they are photons or electrons.
Taking into account figure \ref{relaL}, 
we evaluate the light output for the three background lines (70 keV, 148 keV, and 196 keV). 

% 計算結果
\begin{figure}[htbp]
    \begin{center}
%      \FigureFile(140mm,140mm){figure8.pdf}
      \FigureFile(70mm,70mm){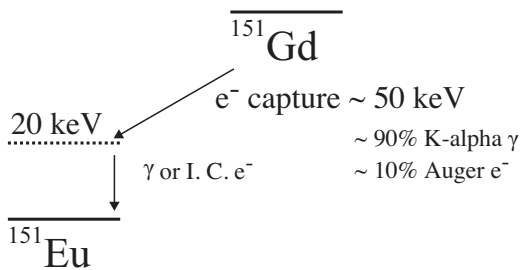}
    \caption{The decay chain from $^{151}$Gd to $^{151}$Eu 
    which produces the background peak in GSO spectra at an energy of 70 keV.}
    \label{decay}
    \end{center}
\end{figure}

For each activation line, we analyzed the decay processes involved therein, 
and calculated their energy releases and branching ratios. 
This was done using such information as nuclear activation energies, 
the energies of fluorescent X-rays, 
the fluorescent yields, and the internal conversion coefficients. 
Then, we calculated light outputs for all the electrons and photons using the corresponding relative light yields in figure \ref{relaL}, 
and summed up the derived light outputs weighted by their branching ratios.  

Figure \ref{decay} shows an example of the decay scheme of the 70 keV line. 
It begins with electron capture of $^{151}$Gd into $^{151}$Eu. 
This releases an energy of $\sim$ 50 keV, via emission of a fluorescent X-ray 
photon or an Auger electron, with a branching ratio of $\sim$ 90\% and $\sim$ 10\%, respectively. 
This leaves $^{151}$Eu in an excited state with a remaining energy of 20 keV, 
which is released by either a photon or an electron through internal conversion. 
A simplified example of evaluation is as follows. 
When a single photon emission with an energy of 70 keV is considered, the light yield is 
estimated as $70 \times 0.92 = 64.4$, where 0.92 can be read 
from figure \ref{relaL}; 
in contrast, where a photon of 20 keV,
plus a 9:1 contribution from a photon of 50 keV 
and an electron of 50 keV are considered, the light yield can be calculated as 
$20 \times 0.78 + 0.9 \times 50 \times 0.9 + 0.1 \times 50 \times 0.96 = 60.9$, 
which is 5.4 \% lower than the single-photon value of 64.4 keV. 
We have calculated more precisely (Yamada 2008), and described them more in detail in Appendix 3.  
As a result, we have found that the light outputs of 70 keV, 148 keV, and 196 keV lines decrease  
by 5.7\%, 5.2\% and 3.0\%, respectively. In other words, for example, 
the activation line which totally deposits 70 keV in GSO should have been regarded 
as a light output from a single photon with an effective energy of 65.8 keV.

\subsection{Re-analysis of the In-orbit Data}

% 光量再計算 figure 2(c)
When the revised output (subsection 4.4) is used, 
the in-orbit data points for the 70, 150, 196 keV lines 
move to the left in figure \ref{newplot}a. 
Then, their ratio to the proportional line have become as shown in figure \ref{newplot}c. 
Thus, the in-orbit data agrees better with the ground data. 
However, the ratio of the in-orbit data still shows 
slightly lower values over 50--400 keV than the on-ground ones.

% 光量再計算+ペデスタルの変化 figure 2(c) 
Finally, the change of the pedestal, 
$\Delta H_{\rm{o}} \sim -8$, as found in section 3.2, was considered. 
As a result, 
the low-enegy in-orbit data points moved in figure \ref{newplot}a
slightly upward along the $H - \Delta H$ axis, 
and their ratios, figure \ref{newplot}d, 
became in full agreement with those of the on-ground data. 
We hence conclude that the in-orbit energy scale of GSO is consistent with the on-ground one within an accuracy of $\sim$ 5\%, 
and the apparent change of the in-orbit data 
is caused partially by the differences of HXD-AE power supplies, 
and more fundamentally, 
by the overestimation of light outputs from the activation lines 
employed in the in-orbit calibration. 
The artificial nonlinearity introduced in Paper II and mentioned in section 1 is no longer needed.

As a further confirmation, 
we overlaid the GSO spectra of A0535+26 in red in figure 3. 
The red spectrum agrees with the blue one at $P \sim 50$, 
which means that the energy scale of GSO agrees with that of PIN at $\sim 50$ keV; 
while the red one around $P \sim$ 70--200 is slightly shifted to the left side by $\sim P$ channels $\sim$ keV, 
which is caused by the recalculation of the light yield preformed in section 4.4. 
Thus, we have succeeded in obtaining 
more correct energy scale of the GSO than the previous one, 
without relying on the artificial correction introduced in Paper II.

\section{Application of the revised energy scale of HXD-GSO}

Since we have confirmed that
the hardware did not change except for the slight pedestal shift,
we have created a new energy scale of GSO 
and implemented it in a ftool {\tt hxdpi} in HEADAS 6.9 or later. 
In this section, we described some after effects of this new energy scale, 
and apply it to the spectra of the Crab Nebula.

\subsection{GSO branch width}

\begin{figure}[tbhp]
    \begin{center}
%      \FigureFile(140mm,140mm){figure9.pdf}
      \FigureFile(70mm,70mm){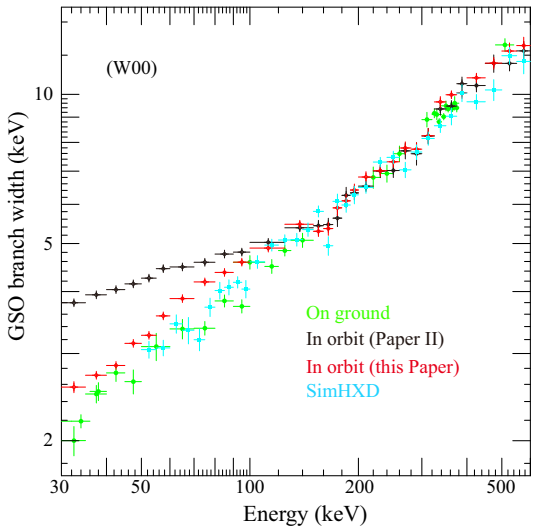}
          \end{center}
    \caption{The GSO branch width (Gaussian sigma) of W00 on the fast-slow diagram, shown as a function of energy. The on-ground and in-orbit data (Paper II) are shown in green and black, respectively. 
    Cyan shows a result of Monte-Calro simulation, and red is in-orbit data based on our new GSO energy scale.}
    \label{psd}
\end{figure}

Let us examine the issue of GSO branch width 
(for definition see figure 15 in Paper II).  
Figure \ref{psd} shows four GSO branch widths from the unit W00 (see caption for details).  
As shown in Paper II and mentioned in section 2.4, 
the in-orbit GSO branch (yellow) became wider below $100$ keV than those of 
the ground calibration (green). 
This change can be explained 
as an effect of the artificial nonlinearity introduced previously,  
because the effect of nonlinearity works differently for fast and slow shaped pulse heights
(unless they happen to have the same value). 
This causes a wider data distribution 
in a fast-slow diagram (cf. figure 2 in Paper II).  

As shown in red in figure 8, 
this problem has nearly disappeared 
when we re-analysed the same in-orbit data 
using the new GSO energy scale established in section 4.5. 
Thus, we have confirmed that the in-orbit GSO branch width is consistent with the on-ground one and that 
the apparent change as explained in section 2.4 was caused by the incorrect energy scale of GSO. 
The revised GSO branch width is released as {\tt ae\_hxd\_gsopsd\_20090812.fits}, which must be 
used in {\tt hxdgrade}. 

\section{Update of GSO response}

The final task of our re-calibtation is to update the response matrix of GSO. 
The calculation of the HXD energy response employes 
a Monte-Carlo simulator, called ``simHXD" (Ozaki et al. 2006; Terada et al. 2005), 
based on a Geant4 framework (\cite{Sulkimo2003} and \cite{Allison2006}).  
which has been constructed using the measured detector performance 
and the actual detector geometry. 
This simHXD assumes a bunch of  input photons obeying a flat spectrum,
and produces the corresponding fake HXD event data,
by simulating how each input photon interacts with the HXD-PIN and/or HXD-GSO
and what kind of signals it produces. 
Then, the simulated data are processed in the same way,
using  the same event selection criteria,
as  the actual data are analyzed. 
This gives an acceptance probability (and hence an effective area)
for an input photon  to be judged as a valid HXD-PIN or HXD-GSO event.

In the present paper, we have improved simHXD in the following three points.
(1) The HXD-GSO energy scale was revised (subsection 5.1), while calculation on energy deposits were not changed. 
(2) The GSO branch width was updated referring to the red data in figure 8.
(2) The simulator was further optimized, with respect to its usage of the GSO brach width information,
      and its emulation of the event selections in HXD-DE (Paper I). 

In figure \ref{psd}, we overlaid a result of a improved simHXD (Terada et al. 2005; Ozaki et al. 2006).
The GSO branch width calculated with simHXD should be consistent with the others.  
In fact, they became consistent as shown in figure \ref{psd}. 
Thus, the on-ground, in-orbit and simulated data became all consistent. 
The new response matrices of HXD-GSO, {\tt ae\_hxd\_gsoxi(hx)nom\_20100524.rsp}, 
has been released in May of 2010 \footnote{http://heasarc.gsfc.nasa.gov/docs/suzaku/analysis/gso\_newgain.html}.  

\subsection{The HXD spectra of the Crab Nebula}

\begin{figure*}[htbp]
    \begin{center}
%      \FigureFile(140mm,140mm){figure10.pdf}
      \FigureFile(140mm,140mm){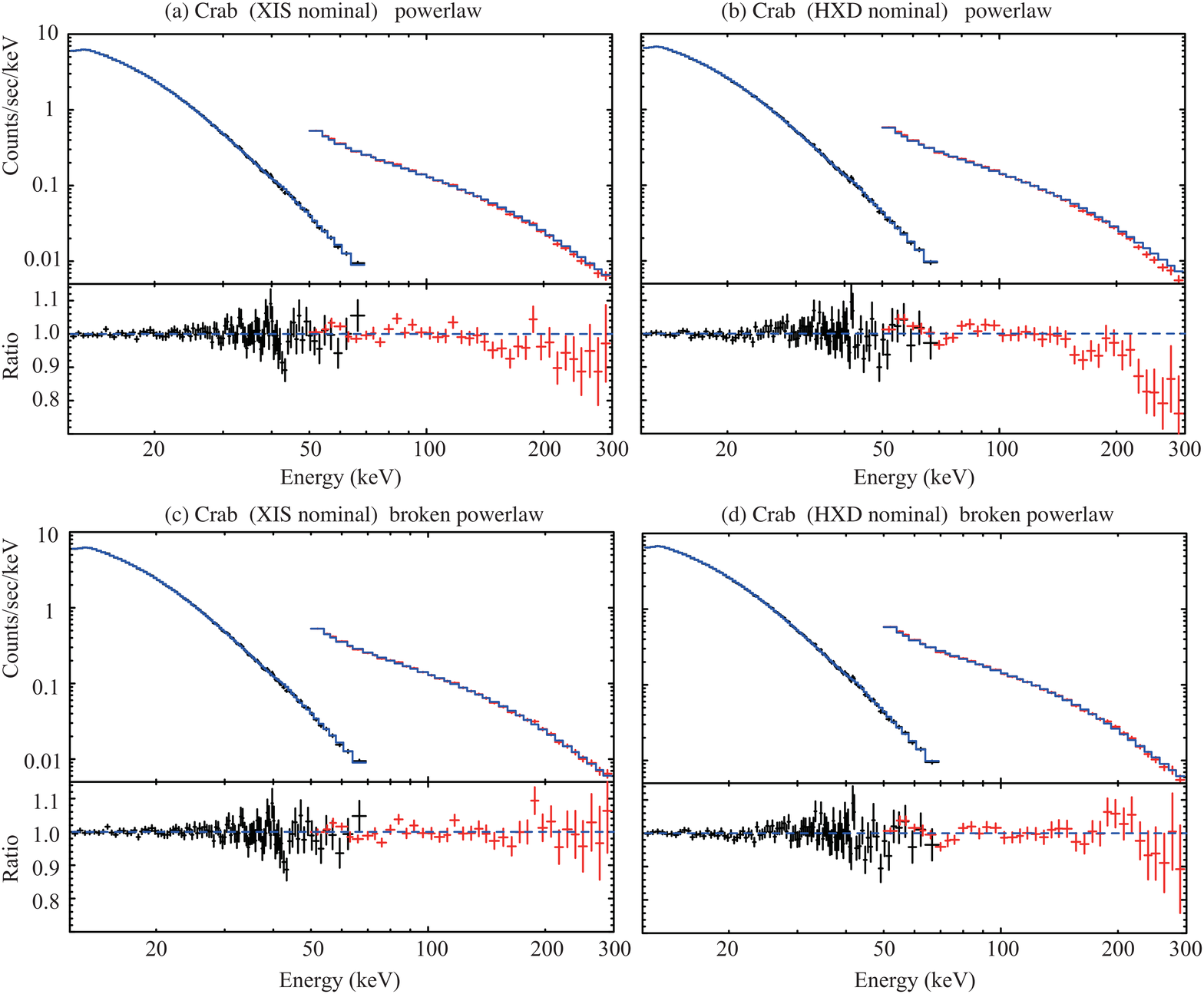}
    \caption{
    The HXD-PIN and HXD-GSO spectra of the Crab Nebula, 
    fitted simultaneously with a power-law model (top two panels labelled a and b) 
   or a broken power-law model (two bottom panels labelled c and d ). 
   }
    \label{crab}
     \end{center}
\end{figure*}

\begin{longtable}{cccccc}
\caption{Fitted parameters with a single and broken powerlaw for the PIN and GSO spectra of the Crab nebula.}
\label{tbl:pin_crabfit}
\hline\hline
Target position &  Photon index1\footnotemark[$^\S$]  & 
Normalization\footnotemark[$*$] & 
Photon index2\footnotemark[$^\S$]  &  
$E_{\rm{break}}$\footnotemark[$^\|$] & 
$\chi_{\nu}^2$(d.o.f)\\[1.05ex]
\endfirsthead
\hline\hline
Target position &  Photon index1 & 
Normalization\footnotemark[$*$] & 
Photon index2 &  
$E_{\rm{break}}$\footnotemark[$^\|$]& 
$\chi_{\nu}^2$(d.o.f)\\
\endhead
\hline
\endfoot
\endlastfoot
\hline
%\hline\hline
%Target position &  Photon index & 
%Normalization\footnotemark[$*$] & $\chi_{\nu}^2$(d.o.f)\\
%\endfirsthead
%\hline\hline
%Target position &  Photon index & 
%Normalization\footnotemark[$*$] & $\chi_{\nu}^2$(d.o.f)\\
%\endhead
%\hline
%\endfoot
%\endlastfoot
%\hline
XIS nominal\footnotemark[$\dagger$] & %  
2.109 $\pm$ 0.003 & 11.65 $\pm$ 0.011 & - & - & 1.23 (132)\\
HXD nominal\footnotemark[$\ddagger$] &  %2005-9-15-14=02
2.096 $\pm$ 0.003 & 11.14 $\pm$ 0.011 & - & - & 1.52 (132)\\[1.3ex]
XIS nominal\footnotemark[$\dagger$]& 
2.104 $\pm$ 0.004 & 11.47 $\pm$ 0.12 &  2.21+0.06-0.04  &  105 $\pm$ 20  &     1.04 (130)\\
HXD nominal\footnotemark[$\ddagger$]& 
2.089 $\pm$ 0.004 & 10.91 $\pm$ 0.12 &  2.28+0.09-0.10 &  117 $\pm$ 19  &     1.09 (130)\\[1mm]
\hline
\multicolumn{4}{l}{\hbox to 0pt{\parbox{180mm}{\footnotesize
\par\noindent
Notes. The column density for the interstellar absorption is fixed at
3$\times$10$^{21}$ cm$^{-2}$.
\par\noindent
\footnotemark[$*$] Power-law normalization in a unit of
photons cm$^{-2}$ s$^{-1}$ keV$^{-1}$ at 1 keV.
\par\noindent
\footnotemark[$\dagger$] 
Observation performed on 2005 September 15 19:50--September 16 02:10 (UT)
\par\noindent
\footnotemark[$\ddagger$] 
Observation performed on 2005 September 15 14:00--19:50 (UT).
\par\noindent
\footnotemark[$^\S$] 
The photon index1 refers to $E<E_{\rm{break}}$, while the photon index2 $E>E_{\rm{break}}$. 
\par\noindent
\footnotemark[$^\|$] 
The break energy of a cutoff powerlaw (keV). 
}\hss}}
\end{longtable}

The Crab Nebula has been used as a standard candle in the X-ray range, 
because it is bright and approximately stable. 
Thus, it is essential to examine whether the new GSO response, created above,
can explain the Crab data, without invoking the artificial correction factor which had been 
introduced to supplement the old response.

We analyzed two data sets of the Crab, 
acquired consecutively on 2005 September 15 each for 6 hours: 
one was obtained at the HXD nominal position from  14:00, and the other at the XIS nominal position from 19:50. 
The GSO data were reprocessed by {\tt hxdpi} and {\tt hxdgrade} with the new GSO branch width, {\tt ae\_hxd\_gsopsd\_20090812.fits} (section 5.1). 
Non X-ray backgrounds of GSO created from the same procedure, released as version 2.4, were used.  
We extracted HXD-PIN spectra from cleaned events, 
and subtracted non X-ray background (Fukazawa et al 2009) from it and used standard responses, 
{\tt ae\_hxd\_pinxi(hx)nome1\_20080129.rsp}. 

The discrepancy in normalization between PIN and GSO became $\sim 20$\%, 
which is probably due to differences between actually observed data and simHXD 
on the anti-coincidence efficiency mainly triggered from Gd-K fluorescence emission lines or 
the depth of the depletion layers in PIN. 
The GSO responses around 50-60 keV is hard to calibrate due to 
its poorer energy resolution at these energy ranges 
and the photo-absorption at Gd-L shell that is not negligible at the lower energy ranges 
and makes the absorption process more complex than in higher energy ranges. 
Since these effects do not significantly affect the spectral shape over $\sim$ 70 keV in GSO, 
we created the GSO auxiliary files, {\tt ae\_hxd\_gsoxi(hx)nom\_crab\_20100526.arf}
\footnote{http://www.astro.isas.ac.jp/suzaku/analysis/hxd/gsoarf2/}, 
which trims (by $\sim$ 20\%) the PIN vs. GSO cross normalization, 
and adjust the HXD-GSO response around  Gd-K edge energy (50--60 keV). 
The cross-normalization between HXD-PIN and HXD-GSO was fixed at 1:1, 
and the fitting range of HXD-PIN and HXD-GSO were 12-70 keV and 50-300 keV, respectively.  
The column density for photoelectric absorption was fixed at $3 \times 10^{21}$ cm$^2$. 
The error refers to 90\% significance. 

We started to fit the HXD-PIN and the HXD-GSO spectra with a powerlaw. 
The obtained parameters are summarized in table 1, 
while the spectra with the best-fitted model (blue) 
and the ratio to the model are shown in figure \ref{crab}a and b, respectively. 
The XIS- and HXD-nominal data were both reproduced approximately by a powerlaw, with reduced $\chi^2$
(d.o.f) of 1.23 (132) and 1.52 (132), respectively. 
In both cases, the derived photon index of $\sim 2.1$ 
and the normalization of $\sim$11 photons s$^{-1}$ cm$^{-2}$ keV$^{-1}$ are close to 
those reported previously.   
However, the ratio to the model, as shown in the bottom panel in figure \ref{crab}a and b, falls $\sim$ 10\% 
below unity in energies above $\sim$100 keV. 
This suggests that there is a mild spectral cutoff at $\gtrsim 100$ keV 
in the Crab spectrum, 
as reported in some previous studies including in particular that with INTEGRAL \citep{crabintegral}. 
 
To express the high-energy turn over, 
we fitted the Crab spectra with a broken powerlaw, 
and obtained the results as summarized in table 1 (bottom two rows) and figure 9 (panels c and d).
The fits became significantly improved to reduced $\chi^2$
(d.o.f) of 1.04 (130) and 1.09 (130), for the XIS and HXD nominal data, 
respectively, and the model vs. data discrepancy has been reduced to $<$10\%. 
The results imply that the photon index steepens from $\sim$ 2.1 to 2.2--2.3 across $\sim$ 110 keV. 
These are consistent with those obtained by INTEGRAL \citep{crabintegral}. 
Thus, we confirmed that the new response of GSO can successfully reproduce the Crab spectrum 
in terms of a broken powerlaw, without using the artificial correction factor needed previously.

\section{Conclusion}

We have performed re-calibration of the energy scale 
of the GSO scintillators which covers the harder energy ranges of the Suzaku HXD. 
Through laboratory experiments using flight-spare hardware, 
we found that the pedestal of HXD-AE decreased by 8 ch when the HXD is driven by the spacecraft power supply, 
than the case where a more stable laboratory power supply is used. 
Calculating all light outputs of secondaries from activation lines revealed that 
the energy of calibrators, used to determine the in-flight gain of GSO, 
was previously over-estimated by several percent. 
Taking both effects into account, the in-orbit data have been confirmed 
to be consistent with the pre-launch calibration. 

When the GSO energy scale was thus revised, 
the in-orbit GSO branch width became also consistent with the ground one. 
Taking into account the energy scale thus updated, as well as the GSO branch width, 
the response matrixes have been created.  
Then, the Crab Nebula spectra of HXD-PIN and HXD-GSO, over 12--300 keV,
has been expressed by a broken powerlaw with a break energy of $\sim 110$ keV 
and a photon index of $\sim$ 2.1 below this energy. 

As a result of these works, major issues with the in-orbit calibration of HXD-GSO have been successfully solved. 

\vspace{0.2cm}

The authors would like to express their thanks to all who have contributed to the design, development, operation, and the calibration of the HXD experiment, and all of the Suzaku supporters. 
The present work was supported by Grant-in-Aid for JSPS Fellows.

\appendix

\section{Irreversible Changes in the HXD Hardware}

The heavy mechanical stress during the launch 
can produce cracks in the GSO crystals,
or degrade their optical contact to the PMT entrance window.
However, the energy resolution of GSO,
measured after the launch using activation lines,
agreed very well with that measured on ground (Paper II).
Therefore, the response of the GSO crystals is 
unlikely to have changed significantly.

The PMTs are another critical item subject to the launch vibration.
Since the launch, the PMTs have been working normally, 
with gain changes by $\sim$0.5\% and $\sim$20\% 
on short ($\sim$hours) and long ($\sim$month) time scales, respectively.
Then, these gain changes may have slightly affected the PMT linearity,
because the PMT bleeders use nonlinear electronic parts, 
such as clump diodes and Zener diodes (Paper I). 
However, the effects of these non-linear components
would appear in the highest energy range 
under large signals caused by cosmic rays,
than in the lower energy range as we observed.
Therefore, the PMT performance is considered 
to have remained essentially unchanged.

Electronics parts in the CSA and HXD-AE 
could in principle be affected by the launch stress.
However, it is highly unlikely that the 16 CSAs and 16 amplification chains,
which individually process signals from the 16 Well-counter units,
were damaged in the same manner.
Considering these, 
we conclude that any conceivable launch-related irreversible hardware change
cannot explain away the present issue of the GSO energy-scale change.

\section{Effects of Large Signals in the Space Environment}

The HXD is exposed in-orbit to a high flux of cosmic rays,
so that the PMTs receive frequent large-amplitude signals.
This could affect detailed PMT responses,
because their bleeders employ non-linear components 
as mentioned in the previous subsection.
Furthermore, the response of HXD-AE may change
when frequent large signals are present.
The effects of large signals on PMTs and HXD-AE
were studied on ground,
by irradiating several isotopes to GSO, 
and optical pulses to the PMT at the same time 
(Tanihata 1998; Kawaharada 2002). 
According to the results, 
any non-linear effect did not appear in energy ranges above $\sim$ 80 keV. 
However, the lower energy range has remained yet to be examined.

\begin{figure*}[htbp]
    \begin{center}
%      \FigureFile(140mm,140mm){figure4.pdf}
      \FigureFile(140mm,140mm){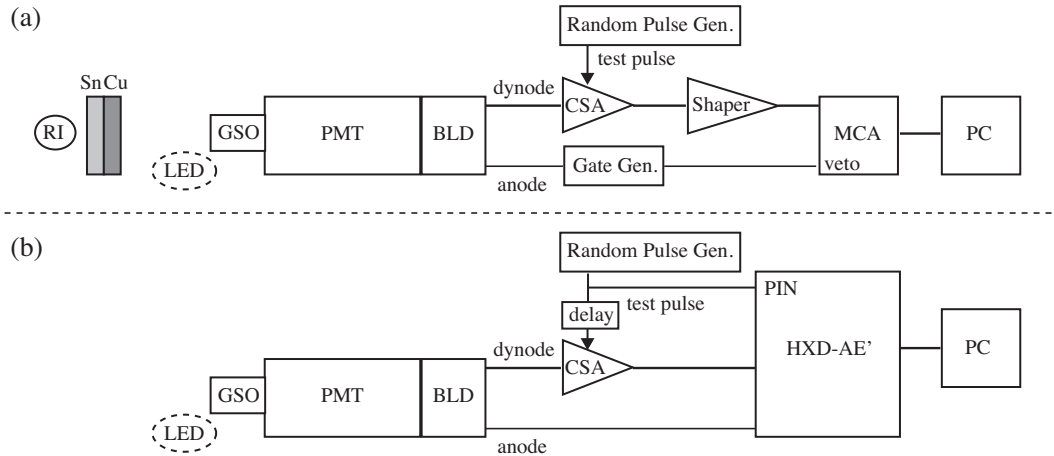}
            \end{center}
    \caption{The experiment setups to test the hardware behavior under large signals. The setup in (a) is used to test 
    the response from GSO to CSA, where that in (b) to test HXD-AE. }
    \label{block}
\end{figure*}

% 1. Flight equivalent hardware  
To study effects of large signals on low-energy events, we performed laboratory experiments 
using a GSO scintillator, a PMT with bleeder, a CSA, and HXD-AE, 
which are all the flight spares or flight equivalent components.  
The flight-spare GSO is of the same size, made by the same process, 
and its energy resolution is several percent worse than those used in the flight hardware. 
% 2. 大信号の起源が何であり、どれくらいの波高値とレートを期待しているかを粛々と述べる。
in-orbit, a Well-counter unit experiences a typical upper-discriminator (UD) hit rate of $\sim$ 100 Hz (Paper II), 
which are mostly caused by primary cosmic rays passing through its BGO shields,  
since the UD threshold is set at $\sim$ 1 MeV. 
Note that a typical minimum-ionizing proton with an energy of $\sim$ 1 GeV deposits an energy of $\sim$ 80 MeV, 
when it passes through the BGO bottom part (with a density of 7 g cm$^{-3}$) for a path length of $\sim 6$ cm. 
%3.  実験のセットアップ
%LED、パルサーとPMTを組み合せて、これをシミュレートするセットアップを説明する。
%大信号の影響を明確にするために、波高値を＊＊倍、レートを＊＊倍おおめに見積もっている。 

Besides the GSO scintillator attached to the PMT, we placed a light-emiting diode (LED) therein, and 
made it flash by a pulse generator to simulate the cosmic-ray signals. 
The LED pulses were adjusted to have a decay time of 500 ns, 
and an energy deposit per pulse of 200 MeV BGO equivalent ( or 50 MeV GSO equivalent) 
with a repetition rate of 100 Hz ($\approx$ the UD hit rate). 
This value is about twice larger than an estimated energy deposit, $\sim$80 keV, 
just to exaggerate possible effects. 
Then, the PMT was operated at 900 V, and the output (last dynode) signal was amplified by the CSA at a time constant 
of 0.5 $\mu$s. The experimental setup is shown in figure \ref{block}.

% 4. Pocket MCA で実験
To confirm the performance from GSO to CSA, 
a commercial multi-chainnel analyzer (Amptek MCA 8000A) was used as an ideal ADC, 
which is known to be linear to within 1 \%.
To test the CSA itself, we input test pulses, created by a random pulse generator at a voltage of 5 V
 (equivalent to $\sim$ 662 keV signals), into the CSA and measure its output using the MCA.    
To avoid the effects of overshoot or undershoot when a large signal (LED signal) comes into the PMT, 
the PMT anode signal was used to veto data acquisition in the MCA via a gate generator. 
As a result, the ADC channels of the test pulses were not affected by the LED signals by more than 1\%.
The result means that the CSA itself does not change under large signals. 

We then proceed to the test from GSO to PMT, using the same setup, 
by irradiating $^{152}$Eu and $^{109}$Cd simultaneously to the GSO scintillator.  
We placed Sn and Cu plates of 5 mm thickness in front of $^{152}$Eu, as shown in gray boxes in figure \ref{block}, 
to stop lower energy ($\lesssim$ 100 keV) events. 
This allowed us to utilize 122 keV and 344 keV lines from $^{152}$Eu, and 22 keV line from  $^{109}$Cd to determine 
the gain and non-linearity of GSO. 
Consequently, the GSO spectrum remained unchanged under the LED irradiation within $<$ 1\%. 
We have thus confirmed that the performance from GSO to CSA, or before HXD-AE, does not change under large signals
% 4. post-FM HXD-AE で実験 

To test the HXD-AE performance under large signals, 
a fight-equivalent HXD-AE system (hereafter HXD-AE') was used, under the setup shown in figure \ref{block}b. 
We put simultaneously into the CSA both the LED signals created by the PMT and test pulses, and measured pulse heights of the test pulse using HXD-AE'. 
Each test pulse from the random pulse generator was fed simultaneously to the CSA (with a 1 $\mu$s delay) 
and ``PIN-trigger'' input to HXD-AE', where the latter was used to acquire the former signal. 
This scheme was necessary to avoid large dead times if the CSA signals were used as self-trigger source.  
Under the presence and absence of the large LED signals, the pulse heights of the random test pulses were 
measured with HXD-AE', for various test-pulse voltages over 0--6 V. 
As a result, the pulse heights measured under the two conditions agreed with $1$ \% without any unexpected effects. 
Consequently, the effect of large signals cannot explain the change of the GSO energy scale.

\section{Calculation of Light Outputs from Activation Lines}

\subsection{Preparation for the Calculation}

Here we show how to calculate light outputs for the 70, 150, 196 keV lines. 
We denote the light outputs of GSO to electron as $L_e(E)$,  
and the one to photon as $L_{\gamma}$.  
To calculate the former, 
we adopted the approximation expression based on the experimental measurement of the light output to electron (Uchiyama et al. 2001), 
\begin{equation} 
L_e (E) =  \left[ 1 - \frac{1.5}{E} \right] \times E.
\end{equation} 
Based on the ground measurement (Kitaguchi et al. 2005), the latter is described as 
\[ E =  E_{\mathrm{LO}}(E) ~  \left[ 1 - \exp ( - 0.058 \times E_{\mathrm{LO}}(E) - 0.53 ) \right].  \]
\begin{equation}
L_{\gamma}(E) = \frac{\left( E_{\mathrm{LO}} (E) - 8.1 ~ \mathrm{keV} \right)}{ \left( E_{\mathrm{LO} } (344 ~ \mathrm{keV}) - 8.1 ~ \mathrm{keV} \right) } \times \mathrm{344}, 
\end{equation} 
where $E_{\mathrm{LO}}$ is an intervening variable. 
The two functions, $L_e (E)/E$ and $L_{\gamma}(E)/E$ are plotted in figure 6. 
The binding energy, the fluorescence energies, and the fluorescence yields for several shells (K, L, M, ..) 
are denoted as $E_{K, L, M, ..}$, $E_{K\alpha, ..}$ and $f_{K, L, ..}$, respectively. 
These parameters of $^{151}$Eu are taken from the database of ENSDF\footnote{http://www.nndc.bnl.gov/ensdf/}.

The light output caused by releasing the energy of the K(L, M, ..)-shell vacancy, 
after the electron bounded in K(L, M..) shell is ejected, should be calculated and defined as $LK(LL, LM, ..)$. 
There are two processes; 
the fluorescence X-ray emitted with the probability of $f_{K(L, M, ..)}$, 
and the Auger electron emitted with the probability of 1 - $f_{K(L, M, ..)}$. 
Considering both effects, 
we can approximately express $LK$ as 
\begin{equation} 
LK  =  f_K \left[ L_{\gamma} (E_{K\alpha}) + L_e (E_K - E_{K\alpha}) \right] + ( 1 - f_K ) L_e (E_K),
\end{equation} 
and $LL$ as 
\begin{equation} 
LL  =  f_L \left[ L_{\gamma} (E_{L\alpha}) + L_e (E_L - E_{L\alpha}) \right] + ( 1 - f_L ) L_e (E_L). 
\end{equation}
In the case of M shell, we assume that all energies are released via the Auger electron, 
expressing $LM$ as  
\begin{equation} 
LM  =  L_e (E_M).
\end{equation}

Then, we define the light output from ejection of the electron from the K(L, M, ..) shell via the internal conversion,  
and denoted them as $L_{eK(eL, eM, ..)}(E)$. It is written as  
\begin{equation} 
L_{eK, eL, eM, ..} (E)  =  L_e ( E - E_{K(L, M, ..)}) + LK(LM, LL, ..), 
\end{equation} 
where $E - E_{K(L, M, ..)}$ represents the kinetic energy of the electron escaping from K(L, M, ..)-shell. 

Finally, 
we define the light output via the gamma decay with the released energy of $E$ as $L_O(E)$. 
$L_O(E)$ becomes a linear combination 
of $L_{\gamma}(E)$ and $L_{eK(eL, eM, ..)} (E)$. 
Each weight of the sum is derived from the internal conversion coefficient for each shell, 
$\alpha$(K, L, M, ..) (ref. ENSDF). 
Therefore, the expression of $L_O(E)$ is 
\begin{equation} 
L_O (E)  = \frac{ \left[ L_{\gamma}(E) + \alpha (K) L_{eK} (E) + \alpha (L) L_{eL} (E) + \alpha (M) L_{eM} (E) ... \right] }{ N }, 
\end{equation}
where the $N$ is the normalization factor and defined as 
\[ N \equiv  1 + \alpha (K) + \alpha (L) + \alpha (M) + ... \]
In the following calculation, we considered the contribution from the electron and gamma emission inside the N-shell.

\subsection{Results of the Light Outputs}

We utilized the internal conversion coefficient of $^{151}$Eu, 
because all the three lines are originated from the beta-decay of Gd.  
The light output of the 196 keV line is a sum of 174 .70 keV and  21.53 keV, 
as described in Paper II.  
By using the notation defined in section A.3.1, 
this can be written as   
\[ 
L_{\gamma} (196.2 ~{\rm keV}) \longrightarrow  L_O (174.7 ~{\rm keV}) + L_O (21.5 ~{\rm keV}) 
\]
According to the equation (A7), 
we obtained $L_O (174.7 {\rm keV})=168.7$ and $L_O (21.5 {\rm keV})=18.3$, 
resulting in 186.9 by summing them, 
while $L_{\gamma} (196.2 {\rm keV})= 192.7$. 
Therefore, the derived light output becomes lowered by 3.0\%. 

Likewise, 
the 150 keV line is the sum of the K-shell binding energy and 
the average energy of 43\% from 103.2 keV and 39.2\% from 97.4 keV, 
which are expressed as  
\[
L_{\gamma} (148.9 ~{\rm keV}) \longrightarrow  \frac{[0.43 L_O (103.2 ~{\rm keV}) + 0.40 L_O (97.4 ~{\rm keV})]}{0.43 + 0.40} + LK.   
\]
We obtained $L_O$ (103.2 keV) = 97.3, $L_O$ (97.4 keV) = 91.4, and $LK$ = 42.3, giving 136.8 in total, 
while $L_{\gamma}$ (148.9 keV) = 144.3.  
Thus, the light output from the 150 keV line becomes lowered by 5.2 \%.
The 70 keV line is the sum of the K shell energy and 21.5 keV, 
expressed by 
\[
L_{\gamma} (70 ~{\rm keV}) \longrightarrow  L_O (21.5 ~{\rm keV}) + LK.   
\]
Considering $L_{\gamma}$ (70 keV) = 64.2, 
the light output from the 70 keV line becomes 60.5, lowered by 5.7 \%.

{}

\end{document}